\documentstyle[aps]{revtex}
\tightenlines
\begin{document}
\rightline{\bf Chin. Phys. Lett. 16 (1999) 256-258}
\begin{center}
\Large Isospin Effects of the Critical Behavior in the Lattice Gas Model \footnotemark \footnotetext{\footnotesize
Supported  by the National Natural Science Foundation for Distinguished
Young Scholar under Grant No. 19725521, the
National Natural Science Foundation under Grant No. 19705012, the
 Science and Technology Development Foundation of Shanghai under Grant No.
97QA14038, the Special Project of  the
Presidential Foundation of Chinese Academy of Sciences,
and the Scientific  Research Foundations for Returned Overseas Chinese
Scholar
by the National Human Resource Administration and Education Administration
of China.}
\end{center}

\begin{center}
\ MA Yu-gang$^{1,2,3}$, \ SU\ Qian-min$^1$, \ SHEN\ Wen-qing$^{1,3}$,
\\ \ WANG Jian-song$^1$, \ CAI\ Xiang-zhou$^1$, \ FANG\ De-qing$^1$

\vspace{.5cm}

$^{1}$Shanghai Institute of Nuclear Research, Chinese Academy of Sciences,
Shanghai 201800, China

$^{2}$Fudan - T. D. Lee Physics Laboratory, Fudan University, Shanghai 200433, China

$^{3}$CCAST (World Laboratory), P. O. Box 8730, Beijing 100080,China
\end{center}
\begin{abstract}
Isospin effects of the critical phenomena were studied via Xe isotopes
in the frame of lattice gas model. All the critical temperatures for 
four Xe isotopes are close to 5.5 MeV at the same freeze-out density of
about 0.39 $\rho_0$. The critical values of
power law parameter of mass distribution, mean multiplicity of 
intermediate mass fragments (IMF), information entropy and
Campi's second moment
show minor dependence on the isospin at the critical point.

\end{abstract}
\vspace{0.5cm}


With the development of the accelerator and radioactive beam technique,
a lot of new experiments using the radioactive beams with large 
neutron or proton excess become possible. The degree of freedom 
of isospin  of nuclear matter is becoming important for research. 
It offers the possibility to study the properties of nuclear matter 
in the range from symmetrical nuclear matter to pure neutron matter.
Some theoretical investigations to the equation of state, chemical 
and mechanical instabilites as well as liquid-gas phase transition for
isospin asymmetrical nuclear matter  were performed already.
In addition, the isospin dependent nucleon - nucleon cross section is 
also an important subject due to its significant effects on the 
dynamical process of heavy ion reactions induced by radioactive beams. 
Some new phenomena stemmed from the isospin have been revealed. For 
examples, the isospin dependences of the preequilibrium nucleon 
emission, the nuclear stopping,  the nuclear collective flow,
total reaction cross section, radii of neutron-rich nuclei
and  subthreshold pion production have been studied by
several groups \cite{Bali98,Macpl,Xucpl}.
However,  more experimental and theoretical
studies are still needed for understanding the isospin physics. As a 
trial, the isospin effects were investigated with the lattice gas
model in this letter.

The lattice gas model of Lee and Yang \cite{Yang52}, in which the 
grancanonical partition function of a gas with one type of 
atoms is mapped into the canonical ensemble of an Ising model for 
spin 1/2 particles, has uccessfully described the liquid-gas
phase transition for atomic system. The same model has already 
been applied to nuclear physics for isospin symmetrical systems 
in the grancanonical ensemble \cite{Biro86} with an approximate 
sampling \cite{Mull97} of the canonical ensemble
\cite{Jpan95,Jpan96,Camp97,Gulm98},
and also for isospin asymmetrical nuclear matter in the mean field 
approximation \cite{Sray97}. In this model, $A$ nucleons with an 
occupation number $s$ which is defined as $s$ = 1 (-1) for a proton 
(neutron) or $s$ = 0 for a vacancy, are placed in the $L$ sites of 
lattice. Nucleons in the nearest neighbouring sites have 
interaction with an energy $\epsilon_{s_i s_j}$. The hamiltonian
is written by 

\begin{equation}
E = \sum_{i=1}^{A} \frac{P_i^2}{2m} - \sum_{i < j} \epsilon_{s_i s_j}s_i s_j
\end{equation}

The interaction constant $\epsilon_{s_i s_j}$ is fixed to reproduce
the binding energy of the nuclei, $\epsilon_{nn,pp}$ = 
 $\epsilon_{-1-1,11}$  = 0. MeV, $\epsilon_{pn,np}$  =
 $\epsilon_{1-1,-11}$ = -5.33 MeV. We use a three-dimension cubic 
lattice L with a size l, a number of nucleons $A = N + Z$ and a
temperature T. The freeze-out density of disassembling system is 
$\rho_f$ = $\frac{A}{L} \rho_0$ where $\rho_0$ is the normal nucleon
density. The disassembly of the system is to be calulated 
at $\rho_f$, beyond which nucleons are too far apart to interact.
$A$ nucleons are put in $L$ cubes by Monte Carlo sampling using the 
Metropolis algorithm \cite{Metr53}. Once the nucleons have been placed, their 
momentum is generated by a Monte Carlo sampling of Maxwell Boltzmann
distribution. Various observables can be calculated in a straightforward
fashion.

One of the basic measurable quantities is the distribution of
fragment mass. In this lattice gas model, two neighboring nucleons 
are viewed to be in the same fragment if their relative kinetic energy 
is insufficient to overcome the attractive bond: 
$P_r^2/2\mu + \epsilon_{s_i s_j} < 0 $. This method is similiar to the so-called
Coniglio-Klein's prescription \cite{Coni80}. In this letter, we use 
the above condition to construct the fragments and their distributions.

We chose several isotopes of Xe as examples of the study of isospin
effects in the lattice gas model. Their isospin parameter
($\frac{N-Z}{A}$) is 0.11, 0.16, 0.21 and 0.26 for $^{122}$Xe,
$^{129}$Xe, $^{137}$Xe
and $^{146}$Xe, respectively. The freeze-out density $\rho_f$
has been chosen to be close to 0.39 $\rho_0$, extracted from the analysis of
Ar + Sc \cite{Jpan95} and 
$^{35}$Cl + Au and $^{70}$Ge + Ti \cite{Beau96} with the 
same model. There is also good support from experiment that the value
of $\rho_f$ is significantly below 0.5$\rho_0$ \cite{Agos96}.  
We use the 343 cubic lattice with size of 7 which results that
the freeze-out density $\rho_f$/$\rho_0$ of $^{122,129,137,146}$Xe is 
0.36, 0.38, 0.40, and 0.43, respectively. The other input 
parameter is the temperature, we perform the calculation from 4 to 7 
MeV. For each isotope 1000 events are accumulated at each temperature. 
 
Fig.1 shows the mass distribution of fragments at T = 4, 5, 6 and 7 MeV
for $^{129}$Xe.
Clearly the disassembling mechanism evolves with the nuclear 
temperature. A few light particles and fragments are emitted and the 
big residue reserves at T = 4 MeV which indicates typical 
evaporation mechanism. 
With the increasing  temperature, the shoulder of mass distribution
occurs due to  the competition between the fragmentation and the evaporation. 
This shoulder disappears and the mass distribution becomes power law shape
at T = 6 MeV,
corresponding to  the multifragmentation region. When the temperature becomes 
much higher, the mass distribution becomes
 steeper indicating that the disassembling process becomes more violent. 
The power law fit, Y(A) $\propto$ $A^{-\tau}$, for these mass distribution
can be  introduced here. It
has already been observed that a minimum of power law parameter 
$\tau_{min}$ exists for most systems if the critical behavior 
takes place. The lines in Fig.1 represent the power law fit. 
Fig.2 displays the several physical quantities 
as a function of temperature for Xe nuclei with the different isospin. 
The minimums of $\tau$ parameters in Fig.2a locate
closely at 5.5 MeV for all the systems, which illustrates  
its minor dependence on the isospin. In  other words, there is a universal
mass distribution regardless of the size of disassembling source when 
the critical phenomenon takes place. However, the $\tau$ parameters show
different values outside the
critical region for nuclei with different isospin, eg., $\tau$
 decreases with isospin when T $>$ 5.5 MeV (multifragmentation 
region). Similiarly, the mean multiplicity of
intermediate mass fragment $N_{IMF}$, defined as the number of 
fragments with 3$\leq$ Z $\leq$16 here, has analogous characters in 
Fig.2b \cite{Maprc,Zhengcpl}. There are the maximums for Xe systems near to 5.5 MeV. When the 
temperature becomes higher,  the larger the source, the higher the $N_{IMF}$.

Fig.2c plots the information entropy H as a function of
temperature for Xe isotopes. The information entropy was
introduced by Shannon   in information theory first\cite{Denb85}.
It is defined as
\begin{equation}
 H = -\sum_{i} {p_i ln(p_i)} ,
\end{equation}
where $p_i$ is the probability having "i" produced particles in each event,
the  sum is taken over all multiplicities of products from the
disassembling system. H reflects the capacity of the information or
the extent of disorder. We introduce this  entropy into the nuclear
disassembly here. As expected, the entropy H reveals the peak close
to 5.5 MeV for all isotopes. These peaks indicate that the opening
of the phase space and the number of the states at the critical
point is the largest. In the other words, the
systems at the critical point have the largest
fluctuation which leads to the largest disorder.
After the critical point, the entropy H increases with the
isospin and/or the source size.

In Fig.2d we give the temperature dependences of Campi's second moment
of the mass distribution \cite{Camp88}, which is defined as
\begin{equation}
S_2 = \frac{ \sum_{i \neq Amax} {A_i^2 \times n_i(A_i)}}{A} ,         
\end{equation}

where
$n_i(A_i)$ is the number of clusters with $A_i$ nucleons and the 
sum excludes the largest cluster $A_{max}$, $A$ is the mass of the system.
At the percolation
point $S_2$ diverges in an infinite system and is at maximum in
a finite system. Fig.2d gives the maximums of $S_2$ around
5.5 MeV for different isotopes, respectively. Again, the critical behavior
occurs in the same temperature as other observables.

In conclusion, the critical behaviors are explored for Xe isotopes 
in the lattice gas model, namely the minimum of power-law parameter $\tau$
of mass distribution, the rise and fall of mean multiplicity of IMF, 
information entropy and  Campi's second moment. 
In a narrow region of 
critical point, the features of the above quantities show no 
dependence on the isospin of the disassembling system. It reflects 
that a universal law exists for the same element in the critical
point. On the contrary, these quantities have isospin dependence at 
the same temperature outside the critical
region.  Noting that the information entropy is introduced into such an 
analysis for the first time, and it seems to be useful for the 
searching of critical phenomena in nuclear physics.
It will be interesting and meaningful to have some experiments
to compare our conclusion in the near future.

We would like to thank Dr. Pan Jicai and Subal Das Gupta for helps and fruitful 
 discussions.  Ma Yu-gang would like
to thank NSFC for the receipt of National Distinguished Young
Investigator Fund.

\begin{center}
Figure Captions
\end{center}
\widetext

\figure{Fig.1: Mass distribution of $^{129}$Xe at T=4, 5, 6 and 7 MeV.
The lines are the power-law fit.
\figure{Fig.2: Critical observales: 
the $\tau$ parameter from the power law fit to mass distribution (a),
the average multiplicity of intermediate mass fragments (b),
the information entropy (c) and the Campi's second moment  (d)
as functions of temperature and isospin.    }


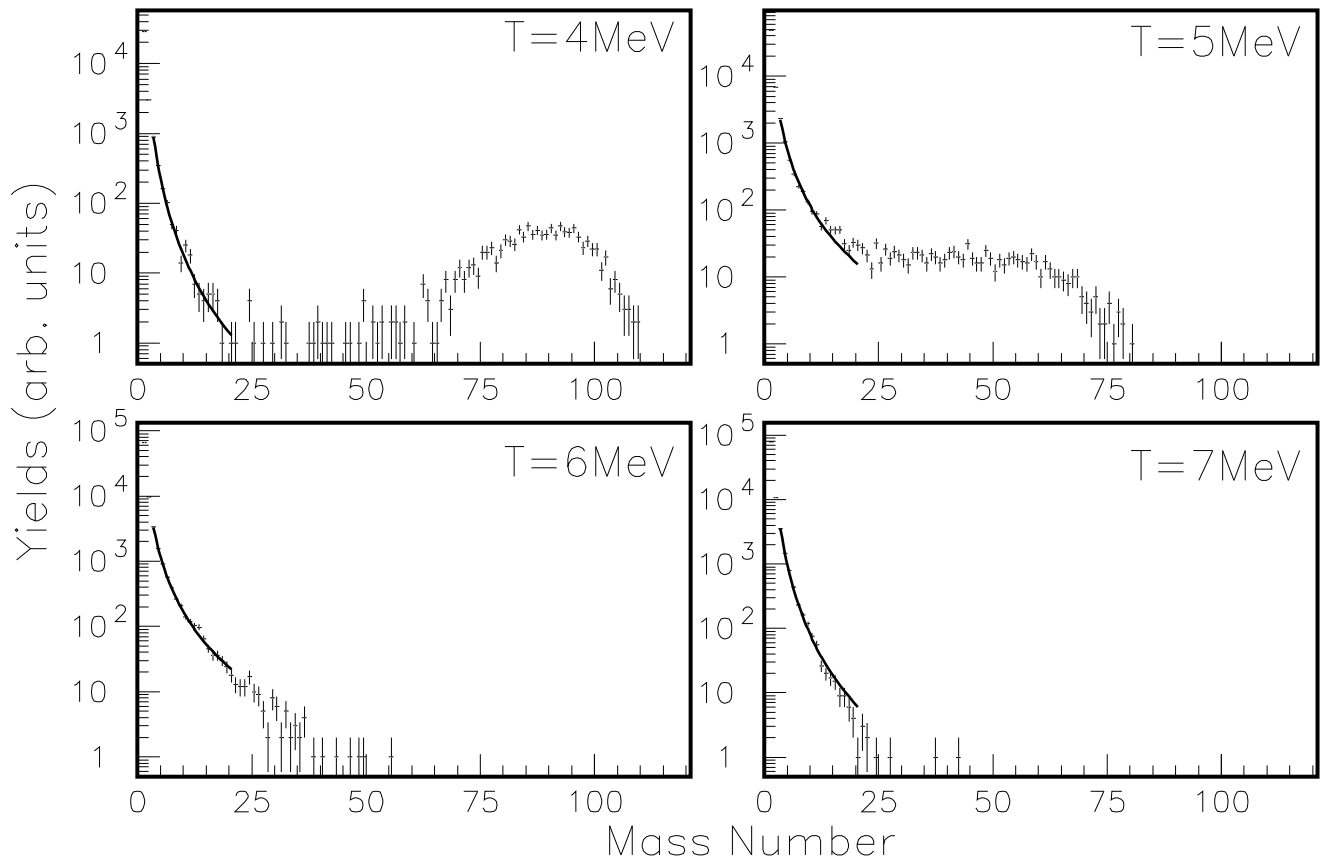
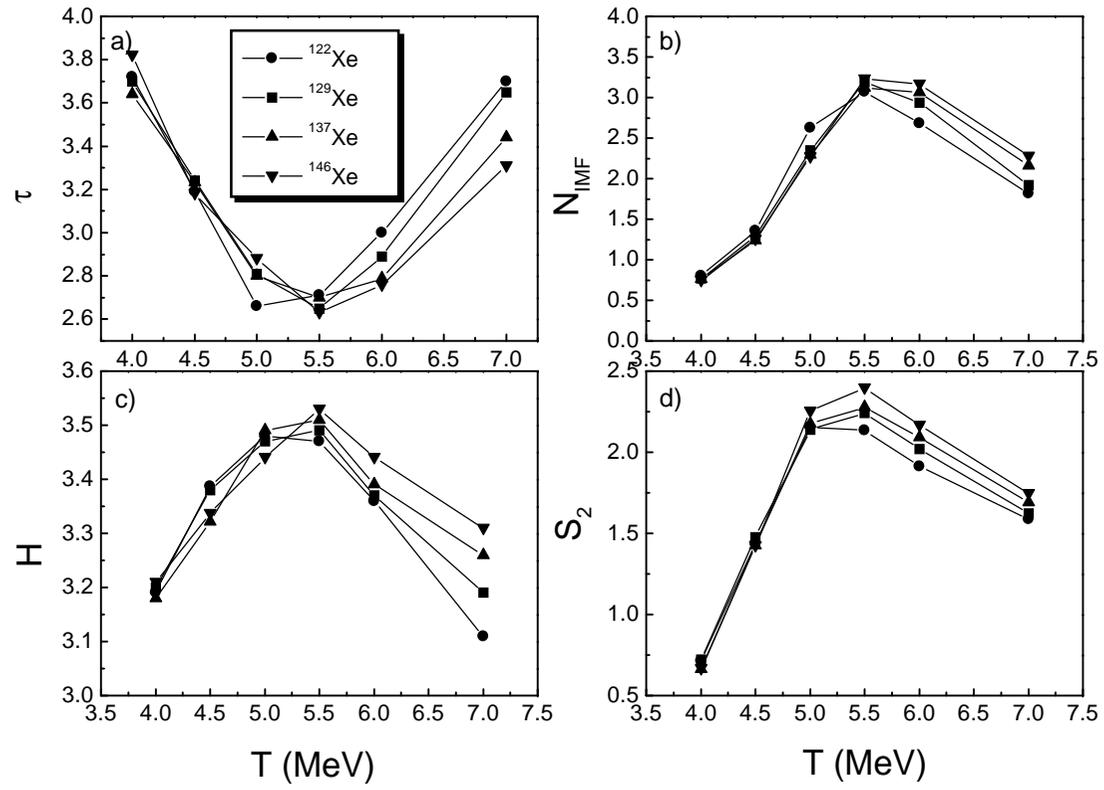
\end{document}